\documentclass{PoS}

\usepackage{amssymb,amsmath}

\usepackage{subfig}

\newcommand{\comments}[1]{}

\newcommand{\fm}{\,\textrm{fm}\,}

\newcommand{\la}{\langle} 
\newcommand{\ra}{\rangle} 

\title{Form factors of the $D \to \pi$ and $D \to K$ semileptonic decays with $N_f = 2$ twisted mass lattice QCD}

\ShortTitle{Form factors of the $D \to \pi$ and $D \to K$ semileptonic decays}

\author{ETM Collaboration}

\author{ \speaker{S.~Di~Vita}$^{ab}$, B.~Haas$^{c}$, V.~Lubicz$^{ab}$, F.~Mescia $^{d}$, S.~Simula $^{b}$ and C.~Tarantino $^{ab}$\\%
  \llap{$^{a}$} Dipartimento di Fisica, Universit\`a Roma Tre, Via della Vasca Navale 84, 00146 Roma, Italy\\
  \llap{$^{b}$} INFN - Sezione Roma Tre, Via della Vasca Navale 84, 00146 Roma, Italy\\
  \llap{$^{c}$} LPT, Universit\'e Paris Sud, Centre d'Orsay, 91405 Orsay-Cedex, France \\
  \llap{$^{d}$} Dep. ECM and ICC, Universitat de Barcelona, Diagonal 647, 08028 Barcelona, Spain\\
}

 \abstract{
 We present lattice results for the vector and scalar form factors of the semileptonic decays \mbox{$D\to\pi\ell\nu_\ell$} and \mbox{$D\to K\ell\nu_\ell$} in the physical range of values of squared four momentum transfer $q^2$, obtained with $N_f=2$ maximally twisted Wilson fermions simulated at three different lattice spacings ($a \simeq 0.102~\fm$, $0.086~\fm$, $0.068~\fm$) with pion masses as light as $270$ MeV and $m_\pi L \gtrsim 4$.
 The form factors are extracted using a double ratios strategy, which allows a good statistical accuracy and is independent of the vector current renormalization constant.
The chiral/continuum extrapolation is performed through a simultaneous fit in the three variables $(m_\pi \,, q^2 \,, a)$ using HMChPT formulae with additional $\mathcal{O}(a^2)$ terms that parametrically account for the lattice spacing dependence.
Our results are in very good agreement with the experimental data in the full $q^2$ range for both $D \to \pi \ell \nu_\ell$ and $D \to K \ell \nu_\ell$. 
At zero momentum transfer we obtain $f^{D\to\pi}(0) = 0.65(6)_{\rm{stat}}(6)_{\rm{syst}}$ and $f^{D\to K}(0) = 0.76(5)_{\rm{stat}}(5)_{\rm{syst}}$, where the systematic error does not include the effects of quenching the strange and the charm quarks. Our findings are in good agreement with recent lattice calculations at $N_f = 2+1$.
}

\FullConference{The XXVIII International Symposium on Lattice Field Theory, Lattice2010\\
		June 14-19, 2010\\
		Villasimius, Italy}

\begin{document}

\section{Overview}
\label{sec:Intro}

 Weak decays of hadrons represent a very important source of direct information about the Cabibbo--Kobayashi--Maskawa (CKM)~\cite{CKM} matrix elements, which are fundamental parameters of the Standard Model (SM) flavour sector.
The direct extraction of such quantities from the experimental decay rates requires theoretical inputs, namely the form factors and the decay constants which encode the non-perturbative QCD dynamics.
Lattice QCD provides a way to compute with a good accuracy these quantities from first principles.

In this contribution we present our determination of the form factors $f_{+,0}^{D \to \pi}(q^2)$ and $f_{+,0}^{D \to K}(q^2)$ obtained using the gauge configurations generated by the European Twisted Mass Collaboration (ETMC) adopting tree-level improved Symanzik gauge action and $N_f=2$ twisted--mass lattice quark action, tuned at maximal twist \cite{ETMC_scaling} to get automatic $\mathcal O(a)$ improvement~\cite{Frezzotti:2003ni}.

The present analysis updates and finalizes our previous study presented in ref.~\cite{ETMC_HeavyLight}, which was based mainly on simulations at a single value of the lattice spacing, $a \simeq 0.086\, \fm$. 
Using the gauge ensembles~\cite{ETMC_scaling} $A_2$--$A_3$ at $\beta = 3.8$ ($a\simeq \, 0.101\, \fm$), $B_2$--$B_4$ and $B_6$--$B_7$ at $\beta = 3.9$ ($a\simeq \,0.086\, \fm$) and $C_2$--$C_3$ at $\beta = 4.05$ ($a\simeq 0.068\, \fm$), we can now extrapolate safely to the continuum limit.
The pion mass ranges from $500$ MeV down to $270$ MeV, and the size $L$ of our lattices guarantees that $m_\pi L \gtrsim 4$, except in the case of the lightest pion for which we have $m_\pi L \gtrsim 3.7$.
For each pion mass and lattice spacing we have simulated several values of the (bare) strange and charm quarks mass to allow for a smooth, local interpolation to the physical values of $m_s$ and $m_c$~\cite{ETMC_masses}.
We impose non-periodic boundary conditions on valence quarks~\cite{Bedaque:2004kc}, which enable us to inject arbitrary values of quark momenta in order to cover the full physical $q^2$ range. 

 We extract the form factors using suitable ratios and double ratios of 2--point and 3--point functions (smoothly interpolated to the physical strange and charm quark masses) at fixed values of $(\,m_\pi,\,q^2,\,a)$. A combined analysis in these variables is then performed in order to reach the physical point. To this end, we fit our data with the predictions of SU(2) HMChPT~\cite{Becirevic}, which describe the mass and momentum dependencies originating from chiral loops in terms of a finite number of LECs, modified by the addition of $\mathcal{O} (a^2)$ terms parametrizing the lattice spacing dependence.

Our preliminary results are in very good agreement with the experimental data in the full $q^2$ range for both $D \to \pi \ell \nu_\ell$ and $D \to K \ell \nu_\ell$.
In particular at zero momentum transfer we obtain $f_+^{D\to\pi}(0) = 0.65(6)_{\rm{stat}}(6)_{\rm{syst}}$ and $f_+^{D\to K}(0) = 0.76(5)_{\rm{stat}}(5)_{\rm{syst}}$, where the systematic error does not include the effects of quenching the strange and the charm quarks. Our findings are in good agreement with recent lattice calculations at  $N_f = 2+1$~\cite{FNAL_MILC_HPQCD}, showing that the error due to the strange quark quenching is smaller than the present uncertainties.

\section{Formalism}
\label{sec:Formalism}

In the Standard Model, CKM matrix unitarity implies that flavour changing decays, at tree level, are due only to charged currents interactions. 
In the case of the semileptonic decay $H_{Qq'}\to P_{qq'} \ell \nu_\ell$, where $H_{Qq'}$ is a \emph{heavy--light} pseudoscalar meson and $P_{qq'}$ is a \emph{light--light} pseudoscalar meson (the subscripts $Qq'$ and $qq'$ represent the valence quark content\footnote{$Q$ is a heavy quark in the sense that $m_Q\gg \Lambda_{QCD}$, while $q$ and $q'$ are light quarks.}) only the vector current contributes to the hadronic matrix element. The decay proceeds at quark level through $Q\to q \ell \nu_\ell$, while the quark $q'$ is just a spectator.
The standard Lorentz decomposition of such matrix element is done in terms of two form factors, $f_+(q^2)$ and $f_0(q^2)$, which encode the non-perturbative QCD dynamics:
  \begin{equation}
    \la P(k)|\bar{q}\gamma_\mu Q|H(p)\ra=f_+(q^2) [ p_\mu+k_\mu-q_\mu (m_H^2-m_P^2)/q^2] +  f_0(q^2) q_\mu(m_H^2-m_P^2)/q^2\,,
  \end{equation}
where $q^\mu = \left(k - p\right)^\mu$ is the four--momentum transfer, $0 \leq q^2 \leq (m_H-m_P)^2$, and the two form factors obey the kinematical constraint $f_+(0)=f_0(0)$.
  
In the case of the semileptonic decay of a heavy meson, another convenient decomposition, in which the form factors are independent of the heavy meson mass in the static limit, is given by
   \begin{equation}
    \label{eq:LorentzDecomposition}
    \la P(k)|\bar{q}\gamma^\mu Q|H(p)\ra=\sqrt{2 m_H} ( v^\mu f_v(E) + p_\perp^\mu f_p(E) )
  \end{equation} 
where $v = p_H/m_H$ is the $H$ meson 4-velocity, $p_\perp = p_P - E v$ and $E = v \cdot p_P = (m_H^2 + m_P^2 - q^2)/(2 m_H)$ is the $P$ meson energy in the $H$ meson rest frame. The two sets of form factors are related by
  \begin{align}
    f_0(q^2) & = \sqrt{2m_H}/(m_H^2-m_P^2)\, \left[ (m_H-E_P)\,f_v(E_P) - p_\perp^2\,f_p(E_P)\right] \,,\\
    f_+(q^2) & = \left[ f_v(E_P) + (m_H-E_P)\,f_p(E_P) \right]/m_H \,.
  \end{align}

The chiral and momentum behaviours of heavy meson form factors are described by the Heavy Meson Chiral Perturbation Theory (HMChPT).
The formulae relevant for our analysis have been computed in continuum SU(2) HMChPT, at next-to-leading order in ChPT and at leading order in the HQET expansion in $1/m_H$ in the partially quenched and unquenched case with degenerate dynamical quarks~\cite{Becirevic}. The explicit formulae for the form factors of the decay $H_{Qq'} \to P_{qq'}$ in our $N_f=2$ lattice setup can be derived from the mentioned paper by computing the unitary limit $m_q = m_{sea}$.
With the addition of a term accounting for discretization effects (starting at $\mathcal{O}(a^2)$ due to automatic $\mathcal{O} (a)$ improvement in maximally tmLQCD), they read\footnote{We label as $m_{xy}$ the mass of a pseudoscalar meson with quark content $xy$.}
  \begin{align}
    f_p(m_{q'q'},m_{qq'},E,a^2) & = \frac{C_0}{E + \Delta} \left( 1 + \delta f_p^{\,(H_{Qq'} \to P_{qq'})} + C_1(E)\, m_{q'q'}^2 + C_2(E) + C_3 \,a^2\right) \,,\\
    f_v(m_{q'q'},m_{qq'},E,a^2) & = D_0 \left( 1 + \delta f_v^{\,(H_{Qq'} \to P_{qq'})}+ D_1(E)\, m_{q'q'}^2 + D_2(E) + D_3 \,a^2\right) \,,
\label{eq:HMChPT_compact}
  \end{align}
where the three contributions represent respectively the non-analytic ($\delta f_{p,v}$) and the analytic terms ($C_{1,2}$, $D_{1,2}$), originating from HMChPT, and the $a^2$ discretization effects ($C_3, ~D_3$). The quantity $\Delta = m_H^* - m_H$ entering the pole factor in $f_p$ is the mass splitting between the vector $H^*$ and $H$. At leading order in the HQET expansion this splitting is zero, and in fact $\Delta$ is consistently neglected in all the loops. Nevertheless, it is customary to keep $\Delta$ in the tree level contribution to $f_p$, since it correctly accounts for the position of the pole expected at $q^2=m_{H^*}^2$ by vector-meson dominance.
For completeness, we collect here the explicit formulae for the non analytic terms:
  \begin{align}
    (4 \pi f)^2 \delta f_p^{\,(H_{Qq'} \to P_{qq'})} = {} & - 9 g^2 I_1(m_{q'q'}) / 4 - I_1(m_{qq'}) + I_1(m_{qq}) / 4 - ( m_{q'q'}^2 - m_{qq}^2)\,\frac{\partial 
    I_1(m_{qq})}{\partial m^2_{qq}} / 4+ \nonumber\\
    & + 6 g^2 J_1(m_{qq'}, E) - 2\, g^2 J_1(m_{qq}, E) + 3\, g^2 \, (m_{q'q'}^2 - m_{qq}^2) \,\frac{\partial J_1(m_{qq}, E)}{\partial m^2_{qq}} / 2 + 
    \nonumber\\
    & + \pi g^2 \, (- 24 m_{qq'}^3 - 9 m_{q'q'}^2 m_{qq} + 17 m_{qq}^3) / 6E \,, \\
    (4 \pi f)^2 \delta f_v^{\,(H_{Qq} \to P_{qq'})} = {} & - 9 g^2 \, I_1(m_{q'q'}) / 4 + I_1(m_{qq'}) + I_1(m_{qq}) / 4 + (m_{q'q'}^2-m_{qq}^2) \frac{\partial 
    I_1(m_{vv})}{\partial m^2_{vv}} / 4 + \nonumber\\
    &  + 2 I_2(m_{qq'}, E) + (m_{q'q'}^2 - m_{qq}^2) \frac{\partial I_2(m_{qq}, E) }{\partial m^2_{qq}} / 2 \,,
  \end{align}
  which in the limit of degenerate valence quarks $q' = q$ read
  \begin{align}
    \label{eq:dp}
    (4 \pi f)^2 \delta f_p^{\,(H_{Qq} \to P_{qq})} = {} & - 3(1+3 g^2) \, I_1(m_{qq}) / 4 + 4 g^2 J_1(m_{qq}, E) - 8 \pi g^2 \, m_{qq}^3/ 3E \,,\\
    (4 \pi f)^2 \delta f_v^{\,(H_{Qq} \to P_{qq})} = {} & (5 - 9 g^2) \, I_1(m_{qq}) / 4 + 2 I_2(m_{qq}, E) \, ,
  \end{align}
where $g$  is related to the coupling constant $g_{D D^* \pi}$. 
The functions $I_1$, $I_2$ and $J_1$ are defined in the $\overline{MS}$ scheme as in Ref.~\cite{Becirevic} and the analytic terms depend on unknown LECs, functions of $E$.
The range of applicability of the NLO chiral logs is generally expected to be limited to energies $E \ll \Lambda_{\rm{ChPT}}$, which implies $q^2$ close to  $q^2_{\rm{max}}$.
However, using the so--called Hard Pion (Heavy Meson) ChPT \cite{HardPion} it has been recently shown that the (HM)ChPT coefficients of the chiral logs are computable also at $q^2=0$, i.e.~far from $q^2 \simeq q_{\rm{max}}^2$. Thus, we will use these chiral predictions to parameterize the mass and momentum dependencies of lattice data in the whole $q^2$ range.

Following  ref.~\cite{FNAL_MILC_HPQCD}, we are also working on a combined extrapolation based on the so-called z-expansion. The outcome of such analysis will be included in a forthcoming publication.
  
\section{Analysis}
\label{sec:Analysis}

 The form factors are extracted from the lattice three-point correlation functions using a double ratios strategy~\cite{Double_Ratios_Damir,Double_Ratios}, which allows a good statistical accuracy and is independent of the vector current renormalization constant, namely
\begin{align}
  \label{eq:double_ratios}
  \frac{C_4^{HVP}\big(\vec{0},t\big) \, C_4^{PVH}\big(\vec{0},t\big)}{C_4^{PVP}\big(\vec{0},t\big)\, C_4^{HVH}\big(\vec{0},t\big)} & \xrightarrow{\rm plateau} R^\prime_0(q_{\rm{max}}^2) \;, \\
  \frac{ C_4^{PVH} \big(\vec{q},t\big)  }{C_4^{PVH} \big(\vec{0},t\big)} \times \frac{C^{PP} \big(\vec{0},t\big)\, C^{HH}\big(\vec{0},t - T/2 \big)}{C^{PP} \big(\vec{q},t\big)\, C^{HH}\big(\vec{q},t - T/2} \big)  & \xrightarrow{\rm plateau} R^\prime_1\big(q^2\big) \;, \\
  \frac{C_i^{PVH} \big(\vec{q},t\big)  }{C_4^{PVH}\big(\vec{q},t\big)} & \xrightarrow{\rm plateau} R^\prime_2\big(q^2\big)\;,
\end{align}
where the ratios $R_i^\prime$ are linear combinations of the form factors $f_{+,0}(q^2)$.

The chiral/continuum extrapolation is performed through a combined $(m_\pi \,, q^2 \,, a)$ fit using the modified HMChPT formulae \eqref{eq:HMChPT_compact} with a polynomial Ansatz for the unknown energy dependence of the analytic terms. We include in the fits the lattice data up to $E \approx 1\, \rm{GeV}$ and we allow terms up to order $E^3$ in the expansion of $C_i(E)$ and $D_i(E)$. 
For the quantity $g$ we adopt the most recent value $g=0.67(14)$~\cite{gDD*pi}, obtained by $N_f=2$ lattice simulations at fine lattice spacings, while the mass splitting $\Delta$ is fixed to the PDG~\cite{PDG} values, $\Delta(D\to \pi)=145\,\rm{MeV}$ and $\Delta(D\to K)=248\,\rm{MeV}$. Finally, for the parameter $f$ we choose the pion decay constant in the chiral limit $f_0=0.1215(1)(^{+1.1}_{-0.1})\, \rm{MeV}$, as determined by the ETM collaboration~\cite{ETMC_scaling}.

\begin{figure}
  \centering
  \subfloat[]{\includegraphics[width=0.47\textwidth]{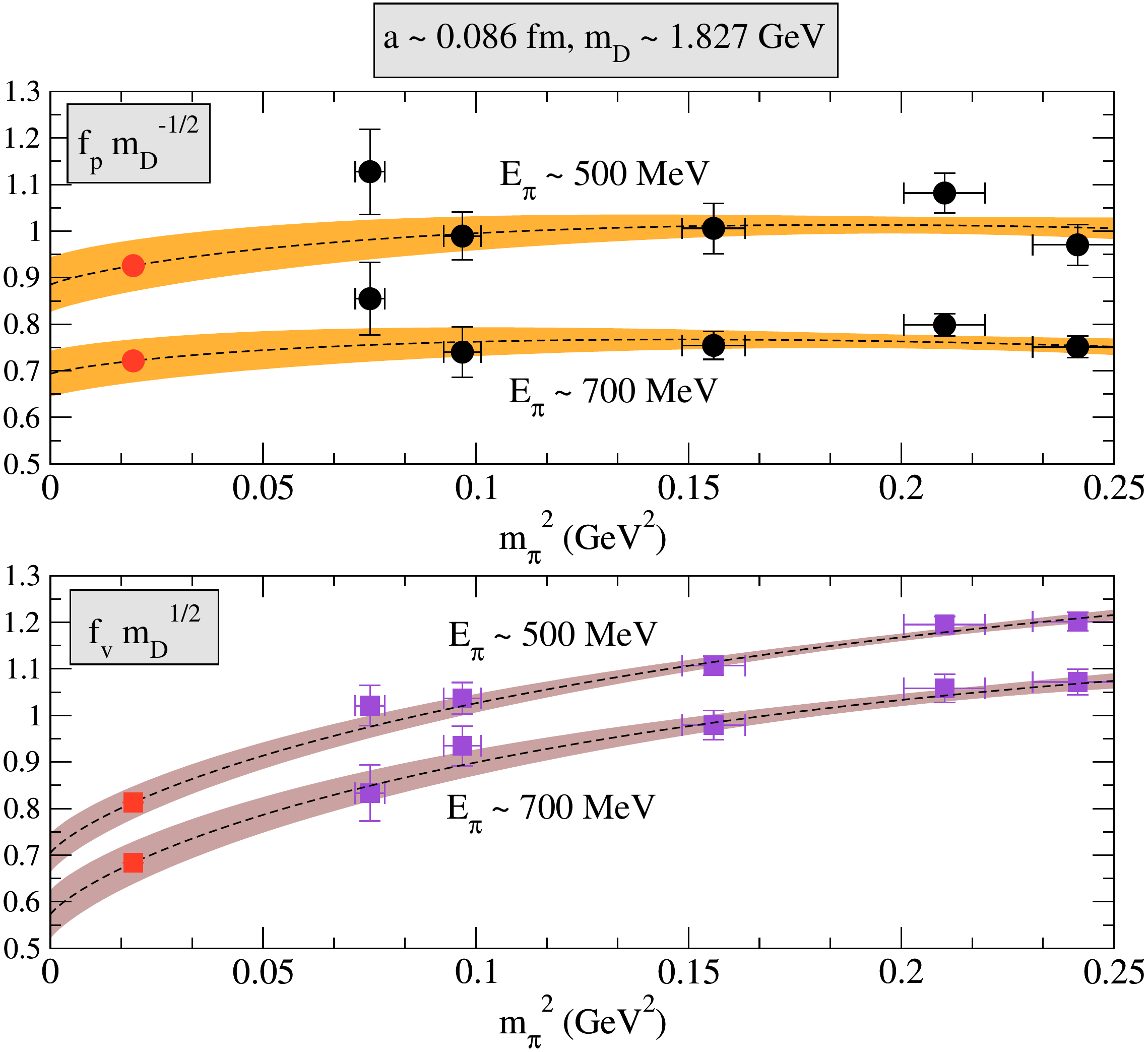}\label{fig:latvsdata_m}} \hfill
  \subfloat[]{\includegraphics[width=0.47\textwidth]{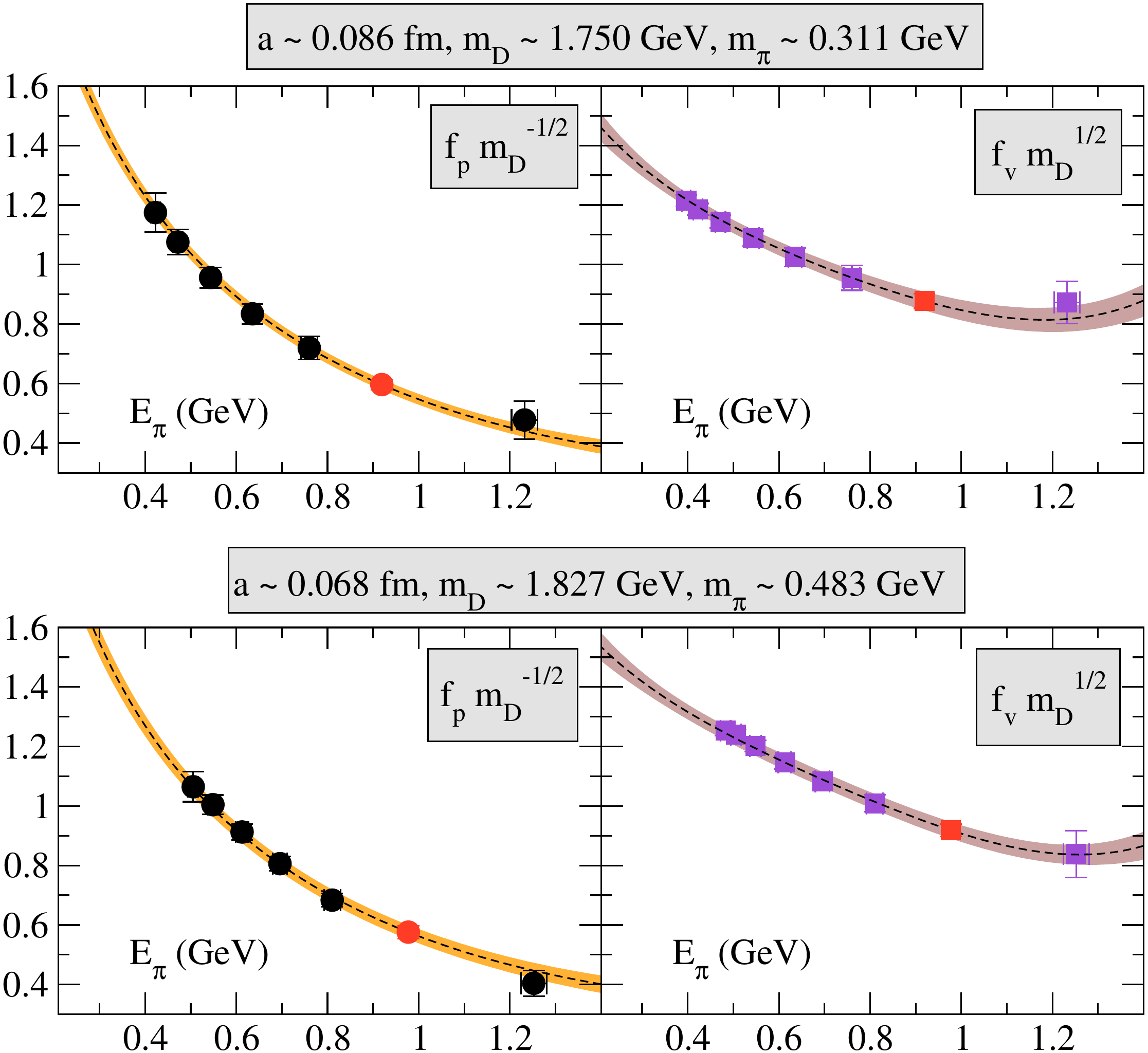}\label{fig:latvsdata_E}} \vfill
  \subfloat[]{\includegraphics[width=0.47\textwidth]{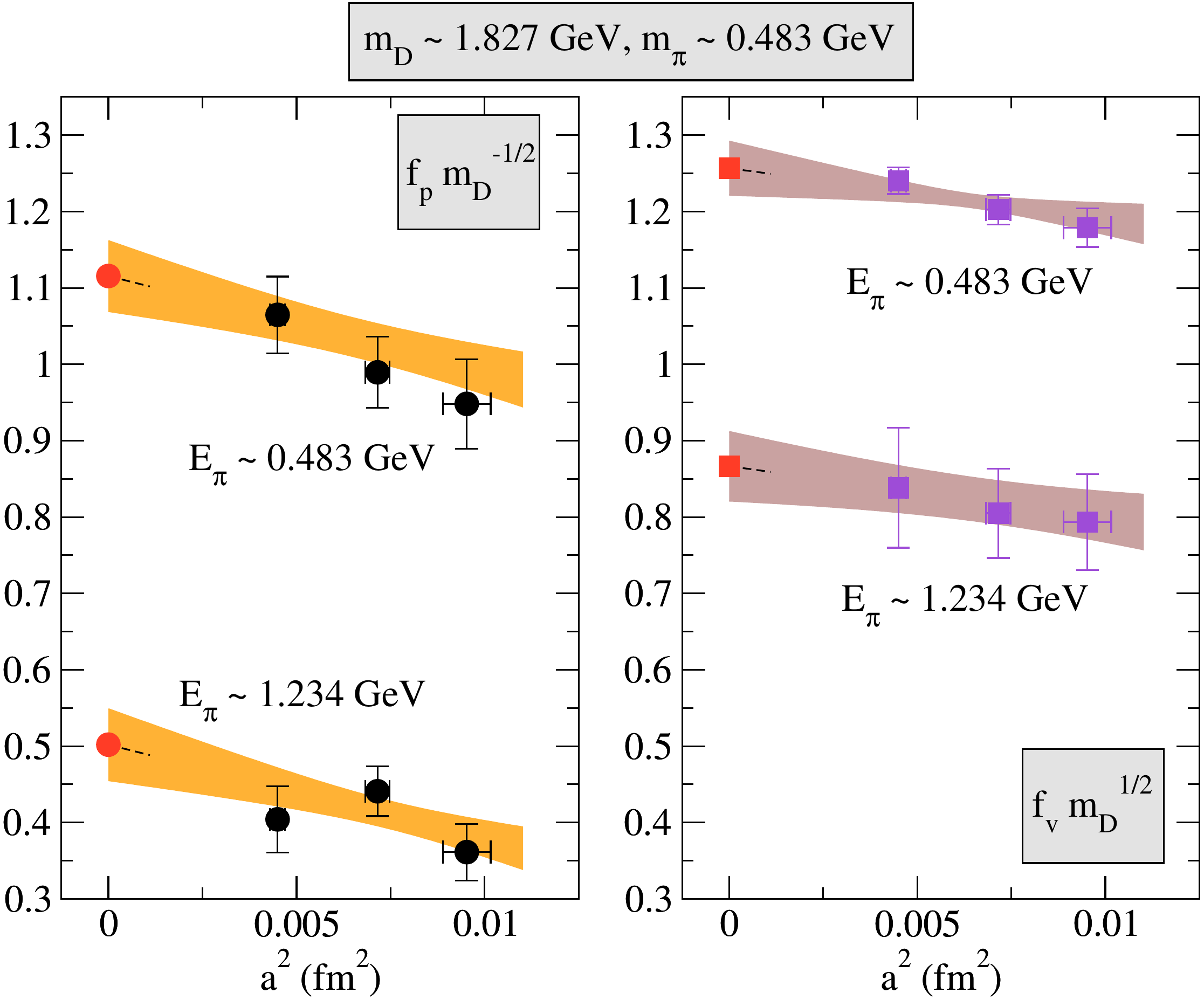}\label{fig:latvsdata_a}} \hfill
  \subfloat[]{\includegraphics[width=0.47\textwidth]{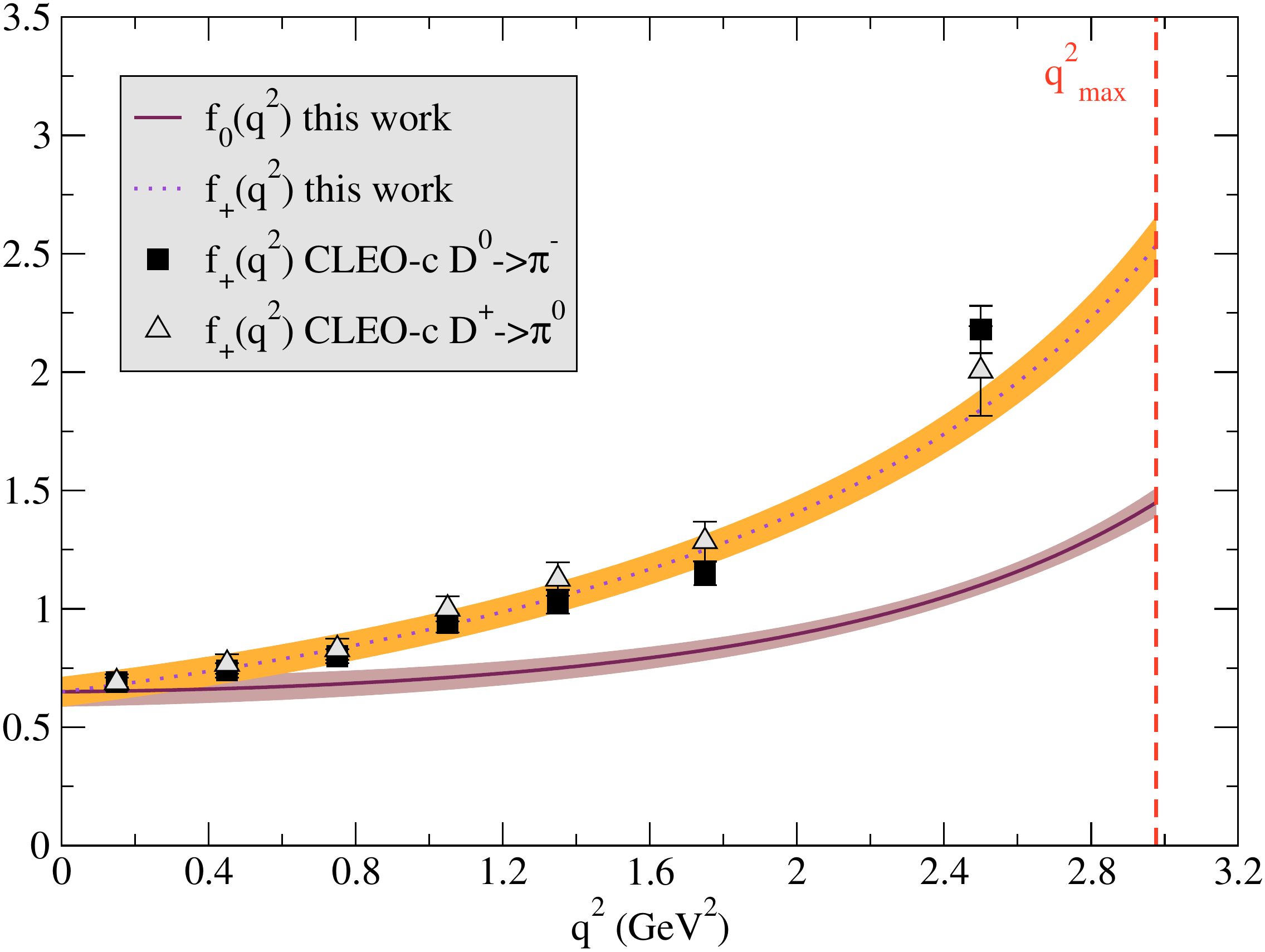}\label{fig:DP_final}}
  \caption[]{\it Pion mass (a), energy (b) and lattice spacing (c)  dependencies of the combined fit for $f_{p,v}^{D\to\pi}$ together with the statistical uncertainty band. The red points show the values of our fit, respectively, at the physical $m_\pi$, $E$ corresponding to $q^2=0$ and $a=0$, while keeping the other two variables fixed. In (d) the results for $f_{+,0}^{D\to\pi}(q^2)$ extrapolated to the physical point (together with the statistical error bands) are shown and compared with the  CLEO-c experimental points~\cite{CLEO-c}.}
\end{figure}

The quality of our fit is illustrated in Figs.~\ref{fig:latvsdata_m}-\ref{fig:latvsdata_a} for the $D\to\pi$ decay. 
Note that the discretization effects appear well described by terms of $\mathcal{O}(a^2)$, as expected. No appreciable dependence of the discretization effects on $m_\pi$ and $E$ is observed.

\begin{figure}
  \centering
  \subfloat[]{\includegraphics[width=0.47\textwidth]{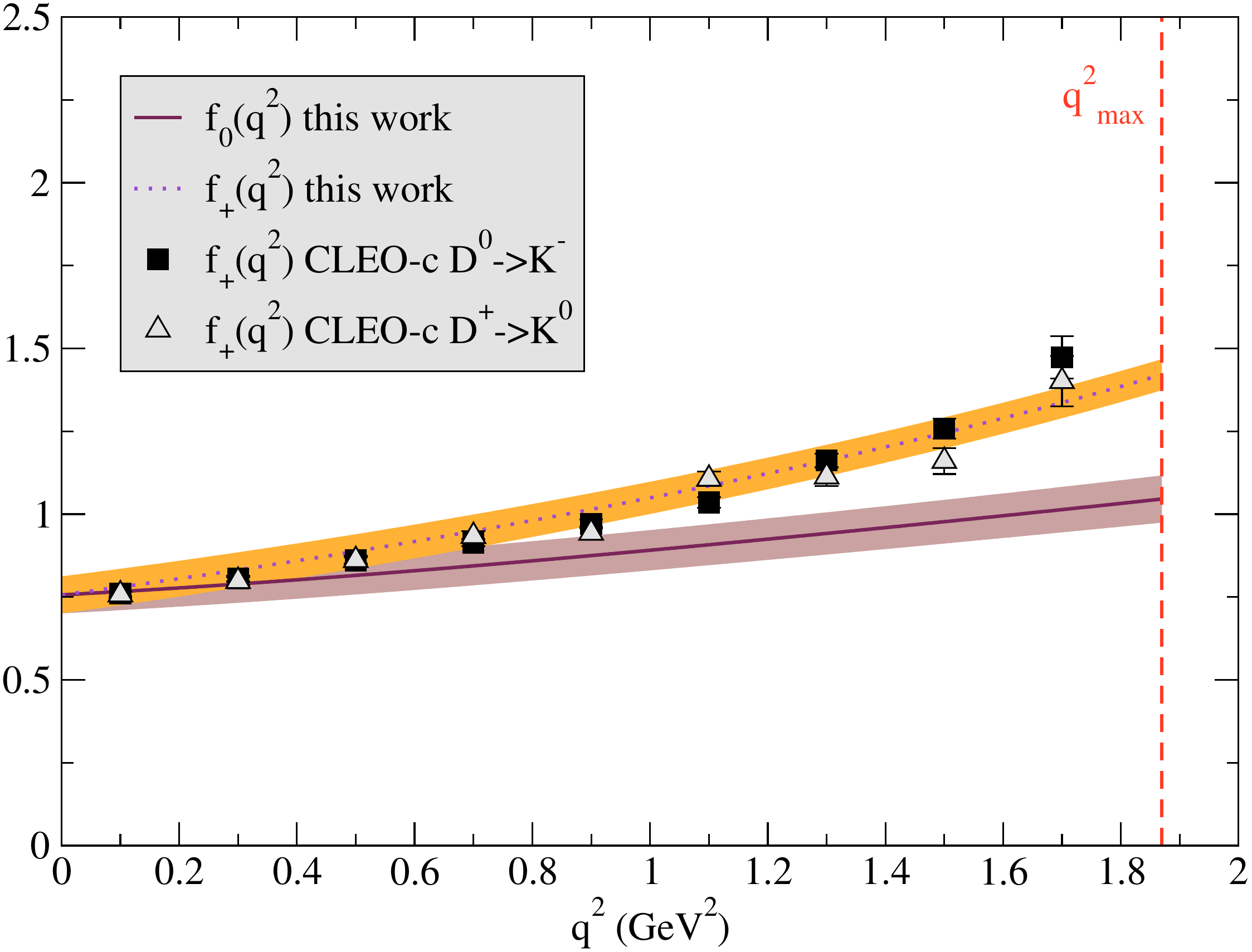}  \label{fig:DK_final}} \hfill
  \subfloat[]{\includegraphics[width=0.47\textwidth]{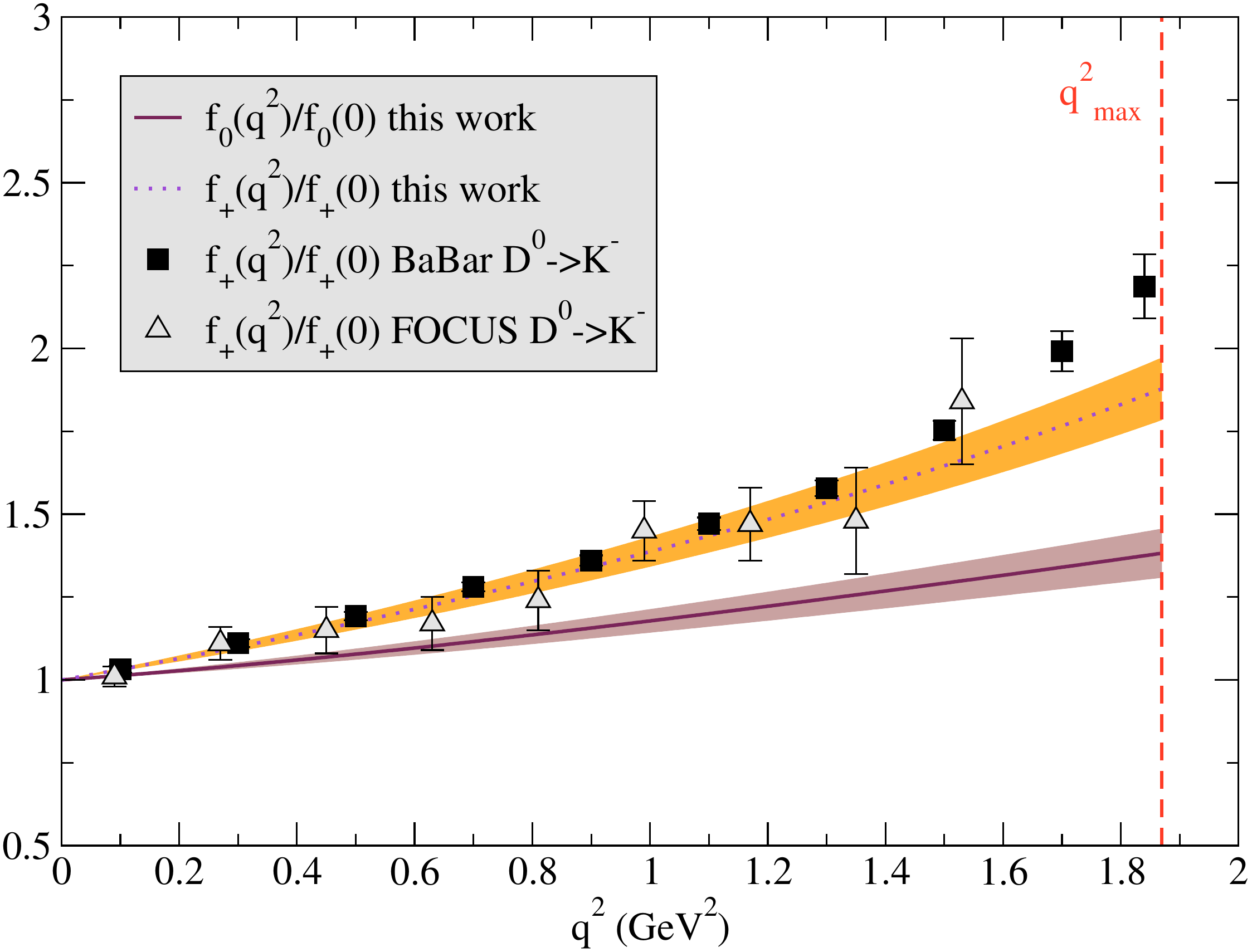}  \label{fig:DK_final_normalized}}
\caption[]{\it Results for the $f_{+,0}^{D\to K}(q^2)$ form factors at the physical point versus $q^2$ (a), compared to the CLEO-c measurements~\cite{CLEO-c}. In (b) the same quantities divided by $f_+^{D\to K}(0)=f_0^{D\to K}(0)$ are compared to the experimental data of BABAR~\cite{BABAR} and FOCUS~\cite{FOCUS} experiments. The bands represent the statistical error of the extrapolation.}
\end{figure}

In Fig.~\ref{fig:DP_final} and Figs.~\ref{fig:DK_final}-\ref{fig:DK_final_normalized} we show the extrapolation of our lattice results to the physical value of $m_\pi$ and in the continuum limit in terms of $f_+(q^2)$ and $f_0(q^2)$. The bands represent the statistical uncertainty resulting from the analysis.

\subsection{Systematic errors}
The fitting procedure is subject to a number of sources of systematic uncertainties, which we are going to discuss in this section except for the effects of quenching the strange and the charm quarks.
We present now a preliminary estimate of the systematic error at zero-momentum transfer, while our final estimate will be included in a forthcoming publication. 

\vspace{0.25cm}
{\it Fitting function and energy range.} We test the stability of our fits by adding/removing terms up to $E^5$ in the LEC's and/or possible NNLO corrections of order $\mathcal{O}(m_\pi^4)$, as well as by including/excluding data with $E_\pi \gtrsim 1 \rm{GeV}$. Terms proportional to $m_\pi E$ and $E^3$ do not modify the fit of $f_p$, which is essentially dominated by the pole factor, while they are necessary for describing the behaviour of $f_v$. We estimate an overall uncertainty of $7\%~(5\%) $ for $D \to \pi$ ($D \to K$) decays.

{\it Discretization effects.} Taking the difference between the result in the continuum limit and at our finest lattice spacing, we estimate that discretization errors are of the order of $5\%~(3\%)$.

{\it Finite size effects.} In the fitting procedure we include the lattice points with the lightest pion, $m_\pi \simeq 270 \textrm{MeV}$, having $m_\pi L \approx 3.7$. By excluding these data, we estimate that volume effects does not exceed $2\%~(2\%)$. 

{\it Value of the coupling constant $g$.} We vary the value of the parameter $g$ in the range $[0.50, 0.67]$ corresponding to available results in the literature. The uncertainty is below $3\%~(3\%)$.

{\it Value of the mass splitting $\Delta$.} We also try to use the lattice determined value instead of the PDG one. This choice increases the statistical error as expected, since the vector meson mass is poorly determined on the lattice, and it does not generate substantial variations in the central values. If treated as a free parameter, it is poorly determined. The estimated uncertainty is $1\%~(1\%)$.

\vspace{0.25cm}
\indent In conclusion, at zero-momentum transfer we get the results
 \begin{align}
    f^{D\to\pi}(0) = 0.65 ~(6)_{\rm{stat}}~(6)_{\rm{syst}}, \qquad f^{D\to K}(0) = 0.76~(5)_{\rm{stat}}~(5)_{\rm{syst}} \, .
 \end{align}
where the systematic error does not include the effects of quenching the strange and the charm quarks.
Our findings are in good agreement with recent lattice calculations at  $N_f = 2+1$~\cite{FNAL_MILC_HPQCD}, showing that the error due to the strange quark quenching is smaller than the present uncertainties.

\section*{Acknowledgements} The authors gratefully acknowledge D. Becirevic for useful discussions and suggestions.


\begin{thebibliography}{99}

\bibitem{CKM} 
  N.~ Cabibbo, 
  Phys.\ Rev.\ Lett.\  {\bf 10} (1963) 531.
 M. Kobayashi and T. Maskawa,  
 Prog.\ Theor.\ Phys.\ {\bf 49} (1973) 652.



\bibitem{ETMC_scaling}
  R.~Baron {\it et al.}  [ETM Coll.],
  JHEP {\bf 1008} (2010) 097
  [arXiv:0911.5061 [hep-lat]].
  Ph.~Boucaud {\it et al.}  [ETM Coll.],
  Phys.\ Lett.\  B {\bf 650} (2007) 304
  [arXiv:hep-lat/0701012];
  Comput.\ Phys.\ Commun.\  {\bf 179} (2008) 695
  [arXiv:0803.0224 [hep-lat]].


\bibitem{Frezzotti:2003ni}
R.~Frezzotti and G.~C. Rossi, 
JHEP {\bf 08} (2004)  007
[arXiv:hep-lat/0306014].

\bibitem{ETMC_HeavyLight}
  S.~Di~Vita {\it et al.} [ETM Coll.],
  PoS {\bf LAT2009} (2009)  257 [arXiv:0910.4845 [hep-lat]]

\bibitem{ETMC_masses}
  B.~Blossier {\it et al.} [ETM Coll.],
  Phys.\ Rev.\  {\bf D82 } (2010) 114513
  [arXiv:1010.3659 [hep-lat]]
 and PoS {\bf LAT2010} (2010) 239 [arXiv:1011.1862 [hep-lat]]

\bibitem{Bedaque:2004kc}
P.~F. Bedaque, 
Phys. Lett. {\bf B593} (2004)  82
[arXiv:nucl-th/0402051].

\bibitem{Becirevic}
  D.~Becirevic, S.~Prelovsek, and J.~Zupan,
  Phys.\ Rev.\ {\bf D67} (2003) 054010
  [arXiv:hep-lat/0210048];
  Phys.\ Rev.\ {\bf D68} (2003) 074003
  [arXiv:hep-lat/0305001].

\bibitem{FNAL_MILC_HPQCD}
  C.~Aubin {\it et al.} [Fermilab Lattice, MILC and HPQCD Coll.],
  Phys.\ Rev.\ Lett.\  {\bf 94} (2005) 011601
  [hep-ph/0408306].
  J.~A.~Bailey {\it et al.},
  PoS {\bf LAT2010} (2010) 306
  [arXiv:1011.2423 [hep-lat]].
  H.~Na {\it et al.},
  Phys.\ Rev.\  {\bf D82} (2010) 114506
  [arXiv:1008.4562 [hep-lat]].




\bibitem{HardPion}
  J.~M.~Flynn, C.~T.~Sachrajda [RBC and UKQCD Coll.],
  Nucl.\ Phys.\  {\bf B812} (2009) 64
  [arXiv:0809.1229 [hep-ph]].
  J.~Bijnens, I.~Jemos,
  Nucl.\ Phys.\ {\bf B840} (2010) 54,
  [arXiv:1006.1197 [hep-lat]].

\bibitem{Double_Ratios_Damir}
  D.~Becirevic, B.~Haas, F.~Mescia,
  PoS {\bf LAT2007 } (2007)  355
  [arXiv:0710.1741 [hep-lat]].

\bibitem{Double_Ratios}
S.~Hashimoto {\it et al.},
  Phys.\ Rev.\  {\bf D61} (1999) 014502
  [hep-ph/9906376].
   D.~Becirevic {\it et al.},
  Nucl.\ Phys.\  {\bf B705} (2005) 339
  [hep-ph/0403217].
  
\bibitem{gDD*pi}
  D.~Becirevic, B.~Haas,
  [arXiv:0903.2407 [hep-lat]].
   D.~Becirevic, B.~Blossier, E.~Chang, B.~Haas,
  Phys.\ Lett.\  {\bf B679 } (2009)  231-236
  [arXiv:0905.3355 [hep-ph]].

\bibitem{PDG}
  K.~Nakamura {\it et al.} [Particle Data Group],
  J. Phys. {\bf G 37} (2010) 075021.

\bibitem{CLEO-c}
  D.~Besson {\it et al.} [CLEO Coll.],
  Phys.\ Rev.\  {\bf D80} (2009) 032005
  [arXiv:0906.2983 [hep-ex]].

\bibitem{BABAR}
  B.~Aubert {\it et al.} [BABAR Coll.],
  Phys.\ Rev.\  {\bf D76} (2007) 052005
  [arXiv:0704.0020 [hep-ex]].

\bibitem{FOCUS}
  J.~M.~Link {\it et al.} [FOCUS Coll.],
  Phys.\ Lett.\  {\bf B607} (2005)  233-242
  [hep-ex/0410037].

\end{thebibliography}
\end{document}